\documentclass[aps,prl,superscriptaddress,twocolumn]{revtex4-1}
\usepackage{bm} 
\usepackage{graphicx} 
\usepackage{amsmath}
\usepackage{amsthm,stmaryrd,mathtools}
\usepackage{amssymb} 
\usepackage{epstopdf} 
\usepackage{txfonts}
\usepackage{graphicx}
\usepackage{siunitx}
\usepackage{hyperref}
\usepackage{braket}
\usepackage{amsmath}
\usepackage{color}

\begin{document}
\title{Scaling of tripartite entanglement at impurity quantum phase transitions}
\author{Abolfazl Bayat}
\affiliation{Department of Physics and Astronomy, University College London,
	Gower Street, London WC1E 6BT, United Kingdom}
\date{\today}

\pacs{05.30.Rt,03.67.Mn,64.70.Tg}

\begin{abstract}
  The emergence of a diverging length scale in many-body systems at a quantum phase transition implies that total entanglement has to reach its maximum there. In order to fully characterize this, one has to consider multipartite entanglement as, for instance, bipartite entanglement between individual particles fails to signal this effect. However, quantification of multipartite entanglement is very hard and detecting it may not be possible due to the lack of accessibility to all individual particles. For these reasons it will be more sensible to partition the system into relevant subsystems, each containing few to many spins, and study entanglement between those constituents as a coarse-grain picture of multipartite entanglement between individual particles.  In impurity systems, famously exemplified by two-impurity and two-channel Kondo models, it is natural to divide the system into three parts, namely, impurities and the left and right bulks. By exploiting two tripartite entanglement measures, based on negativity, we show that at impurity quantum phase transitions the tripartite entanglement diverges and shows scaling behavior. While the critical exponents are different for each tripartite entanglement measure they both provide very similar critical exponents for the two-impurity and the two-channel Kondo models suggesting that they belong to the same universality class.      
\end{abstract}
\maketitle

\emph{Introduction.--} The intrinsic entanglement in the ground state of many-body systems is a resource for quantum technologies \cite{amico2008entanglement}. In particular, at quantum phase transitions, in which the correlation length diverges, critical many-body systems are expected to reach their maximum \emph{total} entanglement, distributed over all length scales.  Nevertheless, neither the entanglement between nearest neighbor particles \cite{osborne2002entanglement,osterloh2002scaling,huang2014scaling} nor the entanglement between a single particle and the rest of the system \cite{larsson2006single} peak at the critical point. This leads to the conjecture that it is the multipartite entanglement that is maximal at criticality. However, verification of this conjecture faces a big obstacle as quantification of multipartite entanglement is still a challenging problem and can only be \emph{evidenced} via appropriate witness operators \cite{seevinck2008partial,huber2010detection,stasinska2014long} or for the case of pure states through either multipartite generalized global entanglement \cite{de2006multipartite} or fidelity approaches \cite{wang2002threshold,bruss2005multipartite}. Some of these methods have also been used in spin chains  \cite{guhne2005multipartite,guhne2006energy,giampaolo2013genuine,giampaolo2014genuine}.

The above conjecture implies that, in a coarse-grained perspective, a hierarchy of different types of entanglement, i.e. bipartite, tripartite, fourpartite  and so on, have to peak at criticality. The most coarse-grained view is the well-established bipartite entanglement between two complementary blocks, quantified via von Neumann entropy, which shows logarithmic divergence at criticality \cite{holzhey1994geometric,calabrese2004entanglement}. Finer levels of coarse-graining will be tripartite, fourpartite and so on, each with an appropriate partitioning, till eventually we reach the true microscopic multipartite entanglement between individual particles. 	 While, all these coarse-grained entanglements are expected to reflect the maximum multipartite entanglement at the critical point, there has been no systematic study for such hierarchical behavior and many fundamental questions are remained to be answered, such as: does entanglement diverge or remain finite at different levels of coarse-graining? can one detect scaling for such entanglement near criticality?


\begin{figure}[t]
	\centering
	\includegraphics[width=\linewidth]{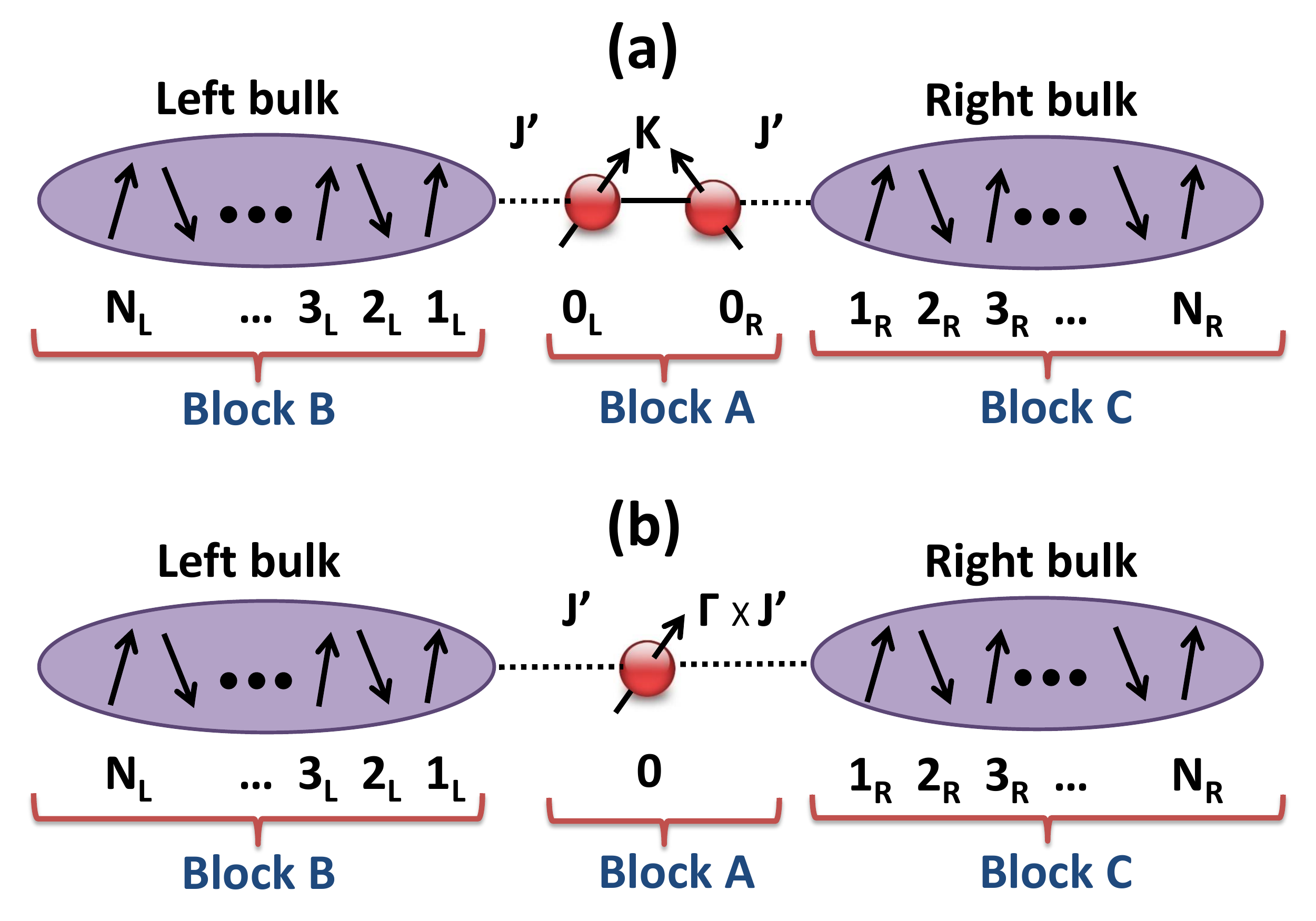}
	\caption{(color online) \textbf{Schematics of the impurity systems.} (a) The 2IKM in which the impurities are coupled to their bulks through impurity coupling $J'$ and interact with each other via RKKY coupling $K$.  (b) the 2CKM in which a single impurity is coupled to two bulks via couplings $J'$ and $\Gamma J'$. In both figures, the system is partitioned into three blocks for studying tripartite entanglement. }
	\label{fig1}	
\end{figure}

Adding one or more impurities to the bulk of a many-body system may change its properties completely leading to new quantum phases \cite{vojta2006impurity}. The impurity quantum phase transitions (iQPTs) cannot be explained by the usual Landau-Ginzburg-Wilson paradigm for bulk quantum phase transitions \cite{amit2005field,sachdev2007quantum} due to the lack of spontaneous symmetry breaking and the absence of local order parameters \cite{bayat2014order,wang2015fidelity}. A typical example for iQPTs arises in  2-impurity Kondo model (2IKM) in which the the Ruderman-Kittel-Kasuya-Yosida (RKKY) interaction between the two impurities competes with the Kondo interaction between each impurity and its bulk. This competition creates a second order quantum phase transition in the 2IKM \cite{bayat2014order,wang2015fidelity}.  
Another crucial model in impurity physics is the 2-channel Kondo model (2CKM) \cite{nozieres1980kondo} in which two independent leads compete  to screen a single spin-1/2 impurity, leading to an ``overscreening" effect \cite{affleck1991critical}. There exist a \emph{critical} crossover, with the emergence of a diverging length scale, at the symmetric case where the two channels equally compete for screening the impurity  \cite{nozieres1980kondo,affleck1991critical,andrei1984solution,affleck1993exact,sengupta1994emery,eggert1992magnetic,mitchell2011real,andrei1995fermi,alkurtass2016entanglement}. 
The 2IKM and the 2CKM are the best examples of non-Fermi liquid behavior generated by criticality \cite{sela2011exact,mitchell2012two,kronmuller2008handbook}. There are also several experimental realizations for both the 2IKM \cite{bork2011tunable,chorley2012tunable,spinelli2015exploring} and the 2CKM \cite{potok2007observation,mebrahtu2013observation,iftikhar2015two,keller2015universal}. 

In this letter, we first introduce two tripartite entanglement measures, which are based on entanglement negativity \cite{lee2000partial,vidal2002computable} for bipartite systems. Then we show that the tripartite entanglement shared between impurities and the two bulks, in both 2IKM and 2CKM, diverges at criticality and shows scaling behavior. Our analysis suggests that the 2IKM and the 2CKM belong to the same universality class.

\emph{Coarse graining.--}  By setting the temperature to zero, we assume that the system is always in its ground state. In structures, such as 2IKM and 2CKM, multipartite entanglement shared between individual spins in the bulk may not be relevant as there might be no access to individual electrons there. Thus, it is more useful to group the particles into certain blocks for which multipartite entanglement can be computed. In both 2IKM and 2CKM a natural partition is to divide the system into three blocks, namely, a block for the impurities and two blocks for the left and the right bulks (see Fig.~\ref{fig1}). While for three qubits there are two independent class of tripartite entanglement, namely the Greenberger-Horne-Zeilinger (GHZ) and the $W$ classes \cite{dur2000three,acin2001classification}, the scenario is far more complicated for many-body systems as such classifications do not exist. 


\emph{Tripartite entanglement.--} Negativity \cite{lee2000partial,vidal2002computable}, as an entanglement measure for bipartite system with density matrix $\rho_{AB}$, is defined as $N_{A,B}=\sum_k |\lambda_k|-1$ where $\lambda_k$'s are the eigenvalues of $\rho_{AB}^{T_A}$ ($\rho_{AB}^{T_B}$) in which $T_A$ ($T_B$) represents the partial transpose  of $\rho_{AB}$ with respect to subsystem $A$ ($B$). Logarithmic negativity, defined as $\log(2N_{A,B}{+}1)$ provides an upper bound for distillable entanglement \cite{plenio2005logarithmic}. Based on negativity, we consider two ways for quantifying tripratite entanglement. The first approach is based on Ref.~\cite{campbell2010multipartite} in which tripartite entanglement is quantified as
\begin{align}\label{Ent1}
E_1=\left[ N_{A,BC} N_{B,AC} N_{C,AB}\right]^{1/3}
\end{align} 
where $N_{A,BC}$ (and equally for the others) stands for negativity between subsystems $A$ and $BC$. This truly quantifies the tripartite entanglement as, for instance, if one subsystem is disentangled from the others then one of the terms in Eq.~(\ref{Ent1}) becomes zero resulting in zero tripartite entanglement no matter whether the two other subsystems are entangled or not. Moreover, since negativity is nonincreasing under local operations \cite{vidal2002computable} the tripartite entanglement $E_1$ will also be the same. It is worth mentioning that for the ground state all the three negativity terms in Eq.~(\ref{Ent1}) are monotonic functions of von Neumann entropies, which uniquely quantify the bipartite entanglement. However, for the sake of generality and consistency with the other measure that will be introduced in the following we use negativity instead of von Neumann entropy.

The second measure for tripartite entanglement is inspired by a generalization of tangle \cite{coffman2000distributed}, as a measure for tripartite entanglement between three qubits. In Ref.~\cite{ou2007monogamy} it was rigorously proved that negativity between three qubits satisfies the inequality $N_{A,BC}^2 \geq N_{A,B}^2+ N_{A,C}^2$. In Ref.~\cite{he2015disentangling} this inequality is conjectured to be valid for arbitrary dimensions based on some numerical investigations and its role for explaining the  robustness of disentangling theorem. Further, numerical analysis confirmed the validity of this inequality in many-body systems \cite{alkurtass2016entanglement}.  Inspired by this inequality the second tripartite entanglement measure is introduced as 
\begin{align}\label{Ent1}
E_2=(\pi_A+\pi_B+\pi_C)/3
\end{align}  
where $\pi_A=N_{A,BC}^2-N_{A,B}^2-N_{A,C}^2$ and similarly $\pi_B$ and $\pi_C$ are determined.


\emph{Model 1: Two impurity Kondo model.--} The first model that we consider is the 2IKM.  The importance of this model lies in the emergence of non-Fermi liquid behavior across its quantum phase transition \cite{sela2011exact,mitchell2012two,kronmuller2008handbook}.
We use the spin chain emulation of the 2IKM \cite{bayat2012entanglement} which is simpler for numerical analysis using Density Matrix Renormalization Group (DMRG) \cite{white1992density}. The Hamiltonian is written as $H{=}\sum_{i=L,R}H_i + H_I$ with
\begin{eqnarray}\label{Ham_2IKM}
H_i&{=}&J'\left(J_1\boldsymbol{\sigma}_0^i\cdot \boldsymbol{\sigma}_1^i + J_2 \boldsymbol{\sigma}_0^i\cdot \boldsymbol{\sigma}_2^i \right)+\cr
&{+}&J_1\sum_{k=1}^{N_i-1} \boldsymbol{\sigma}_k^i\cdot \boldsymbol{\sigma}_{k+1}^i + 
J_2 \sum_{k=1}^{N_i-2} \boldsymbol{\sigma}_k^i\cdot \boldsymbol{\sigma}_{k+2}^i, \cr
H_I&=&J_1K \boldsymbol{\sigma}_0^L\cdot \boldsymbol{\sigma}_{0}^R.
\end{eqnarray}
Here  $i{=}L, R$ labels the left and right chains with $\boldsymbol{\sigma}_k^i$ being the vector of Pauli matrices at site $k$ in chain $i$, and with $J_1$ ($J_2$) nearest- (next-nearest-) neighbor couplings. Impurities sit at site $0$ of each chain and the dimensionless parameters $J'$ and $K$ represent the impurity and RKKY couplings respectively. The total size of the system is $N=N_L+N_R$ and throughout this letter we take $N_L=N_R$. By fine
tuning $J_2/J_1=0.2412$ to the critical point of the spin chain dimerization transition \cite{nomura1994critical,eggert1996numerical}, the Hamiltonian of Eq.~(\ref{Ham_2IKM}) provides a faithful representation of 2IKM \cite{bayat2012entanglement}. The coupling $K$ is the control parameter which we vary by fixing impurity coupling $J'$. For small values of $K\ll J'$, i.e. Kondo phase, each impurity is screened by its own bulk resulting in two independent single impurity Kondo chains. On the other hand for $K\gg J'$, i.e. dimer phase, the two impurities form a singlet and decouple from the system. For some intermediate value of $K=K_c$ a quantum phase transition happens between the two phases which can be detected by Schmidt gap \cite{bayat2014order}. In order to analyze the tripartite entanglement across the quantum phase transition we partition the system into three parts, namely, block $A$ containing the two impurities (i.e. sites $0_L,0_R$), block $B$ containing the spins in the left bulk (i.e. sites $1_L,2_L,\cdots,N_L$) and block $C$ which contains the spins in the right bulk (i.e. sites $1_R,2_R,\cdots,N_R$). A schematic of this is shown in Fig.~\ref{fig1}(a).

\emph{Model 2: Two Channel Kondo model.--} The second system that we consider is the 2CKM \cite{nozieres1980kondo}. 
Similar to the 2IKM and for the sake of simplicity we take the spin chain emulation of the 2CKM \cite{alkurtass2016entanglement} as
$H_{2CK}=\sum_{i=L,R}H_i^{2CK}+H_{int}^{2CK}$ with
\begin{eqnarray}\nonumber
H_i^{2CK}&=& J_1 \sum_{k=1}^{N_i-1} \boldsymbol{\sigma}_k^i\cdot \boldsymbol{\sigma}_{k+1}^i+
J_2 \sum_{k=1}^{N_i-2} \boldsymbol{\sigma}_k^i\cdot \boldsymbol{\sigma}_{k+2}^i \cr
H_{int}^{2CK}&=& J' \left( J_1 \boldsymbol{\sigma}_0 \cdot \boldsymbol{\sigma}_1^L + J_2 \boldsymbol{\sigma}_0 \cdot \boldsymbol{\sigma}_2^L \right) \cr
&+& J' \Gamma \left( J_1 \boldsymbol{\sigma}_0 \cdot \boldsymbol{\sigma}_1^R +
J_2 \boldsymbol{\sigma}_0 \cdot \boldsymbol{\sigma}_2^R\right)
\end{eqnarray}
where $\boldsymbol{\sigma}_0$ represents the impurity spin and $J'$ stands for the impurity coupling with $\Gamma$ being the asymmetry parameter. The total size of the system is $N=N_L+N_R+1$ and throughout this paper we take $N_L=N_R$. In the 2CKM the parameter $\Gamma$ plays the role of the control parameter and the system shows critical behavior around $\Gamma=\Gamma_c=1$, where the two channel Kondo physics is valid, with a diverging length scale $\xi_{2CK}\sim |\Gamma-1|^{-\nu}$. For $\Gamma\ll 1$ (and $\Gamma \gg 1$) system reduces to a single impurity Kondo problem with impurity being screened by the left (right) channel.  In order to study tripartite entanglement we divide the system into three blocks, namely, block $A$ which includes impurity spin (i.e. site $0$), block $B$ which is the left bulk (i.e. $1_L,2_L,\cdots,N_L$) and block $C$ which is the right bulk (i.e. $1_R,2_R,\cdots,N_R$). A schematic of the 2CKM is shown in Fig.~\ref{fig1}(b).

\begin{figure}[t]
	\centering
	\includegraphics[width=\linewidth]{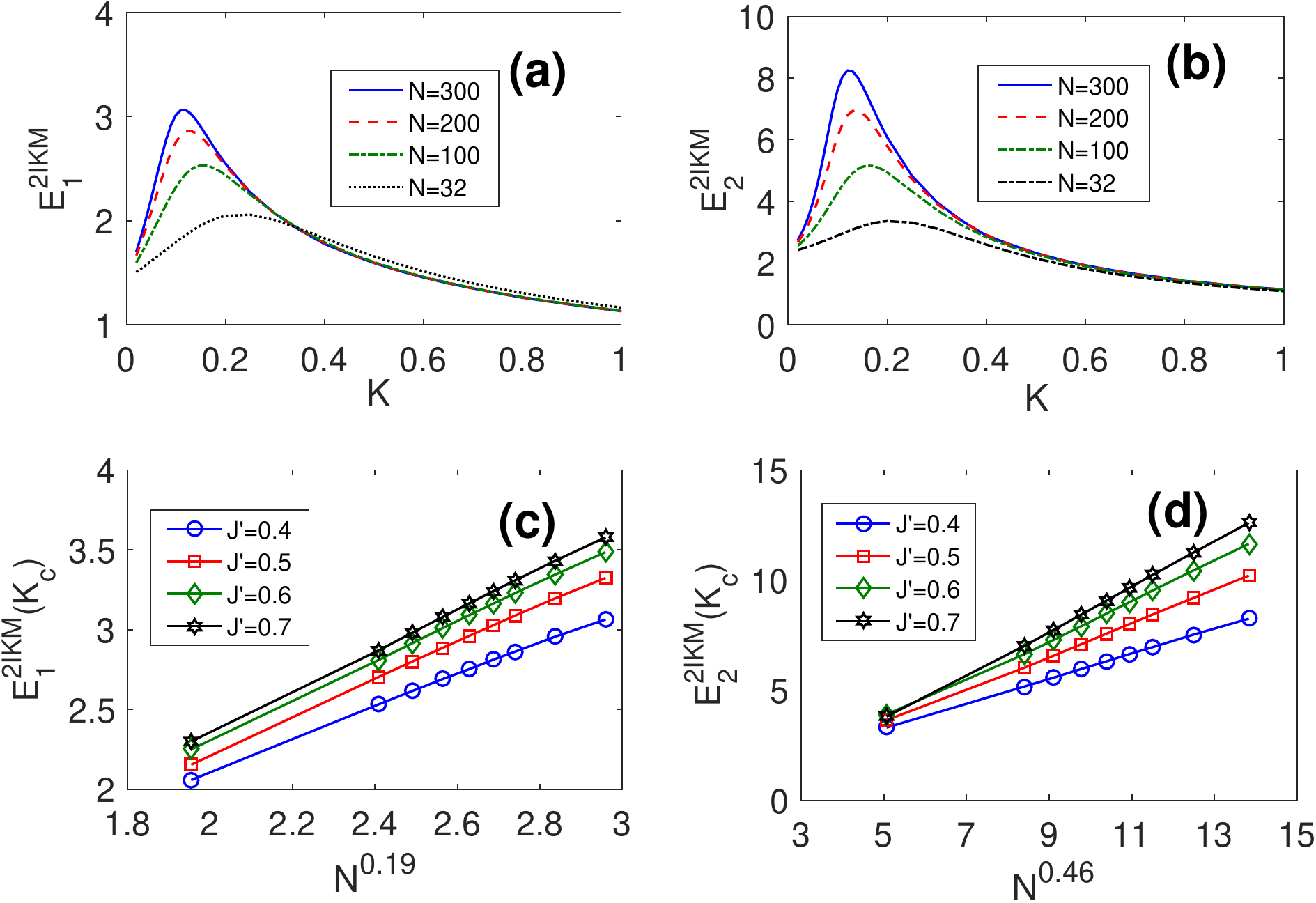}
	\caption{(color online) \textbf{Tripartite entanglement in 2IKM.} (a) Tripartite entanglement $E_1^{2IKM}$ vs. $K$ in a chain with $J'=0.4$. 
		(b) Tripartite entanglement $E_2^{2IKM}$ vs. $K$ in a chain with $J'=0.4$. (c) Scaling of $E_1^{2IKM}(K_c)$ in terms of $N^{0.19}$. 
		(d) Scaling of $E_2^{2IKM}(K_c)$ in terms of $N^{0.46}$.}
	\label{fig2}
\end{figure}

\begin{figure}[t]
	\centering
	\includegraphics[width=\linewidth]{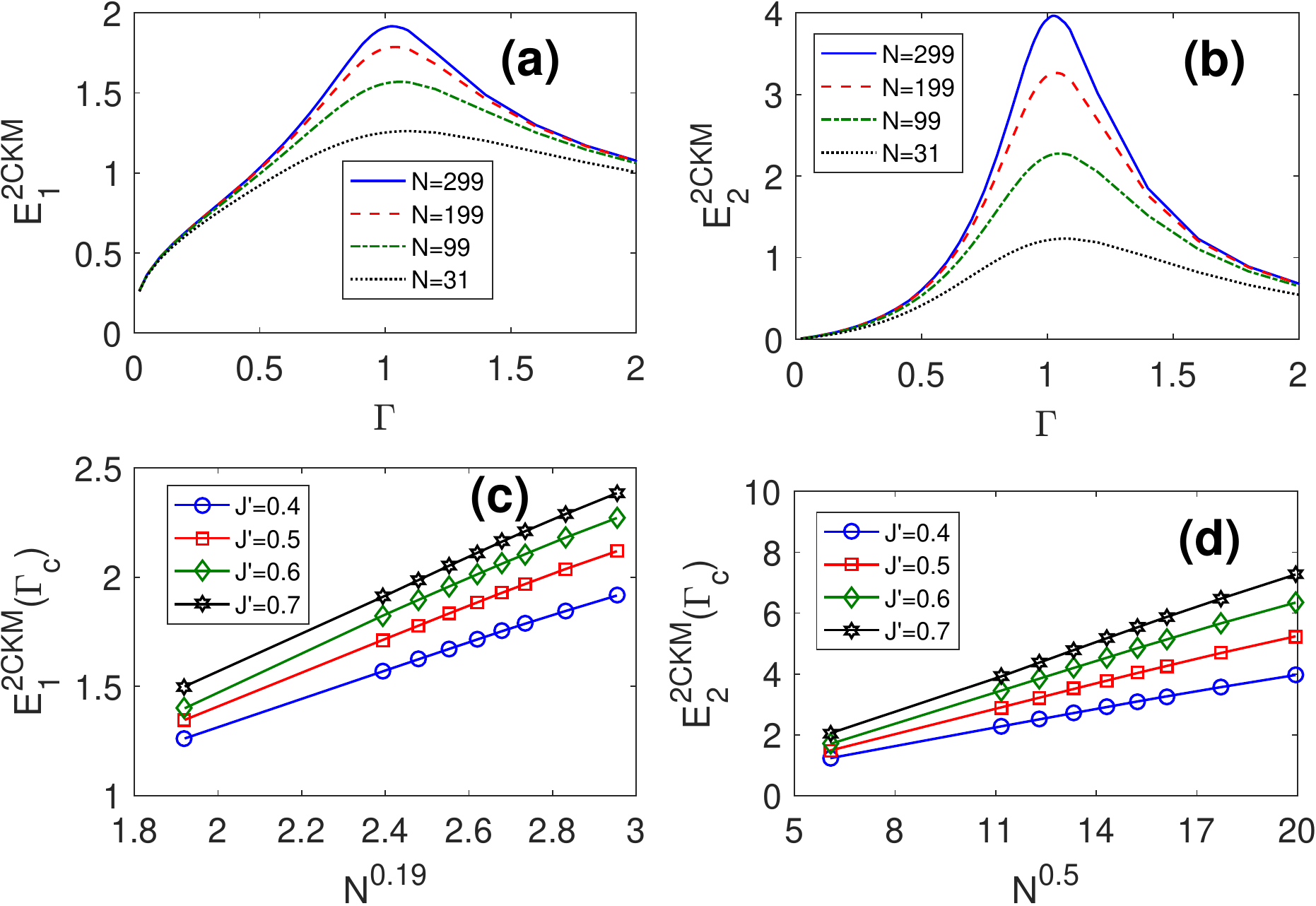}
	\caption{(color online) \textbf{Tripartite entanglement in 2CKM.} (a) Tripartite entanglement $E_1^{2CKM}$ vs. $\Gamma$ in a chain with $J'=0.4$. 
		(b) Tripartite entanglement $E_2^{2CKM}$ vs. $\Gamma$ in a chain with $J'=0.4$. (c) Scaling of $E_1^{2CKM}(\Gamma_c)$ in terms of $N^{0.19}$. 
		(d) Scaling of $E_2^{2CKM}(\Gamma_c)$ in terms of $N^{0.5}$.}
	\label{fig3}
\end{figure}

\emph{Divergence of tripartite entanglement.--}  We study the tripartite entanglement, quantified by both $E_1$ and $E_2$, across the phase diagram of the 2IKM and the 2CKM using DMRG. In the 2IKM the RKKY coupling $K$ and in the 2CKM the asymmetry parameter $\Gamma$ are varied as control parameters for any fixed impurity coupling $J'$. The results for 2IKM are depicted in Figs.~\ref{fig2}(a)-(b) for an impurity coupling $J'=0.4$ and various system sizes. As the figures clearly show both $E_1^{2IKM}$ and $E_2^{2IKM}$ peak at a specific value of $K{=}K_c$ and the peaks become more pronounced by increasing the system size. This suggests that the tripartite entanglement diverges at $K=K_c$ in the thermodynamic limit (i.e. $N \rightarrow \infty$). The critical point $K_c$ is proportional to the Kondo temperature as $K_c\sim e^{-\alpha/J'}$ (data are not shown for this) which is in full agreement with \cite{bayat2014order,bayat2012entanglement}. In order to see how tripartite entanglement diverges we can compute $E_j^{2IKM}(K_c)$ (for $j=1,2$) for various system sizes. One can numerically verify that 
\begin{align} \label{Emax_2IKM}
E_j^{2IKM}(K_c) \sim N^{\lambda_j^{2IKM}} \quad \text{(for } j=1,2), 
\end{align}
where both of the exponents $\lambda_1^{2IKM}$ and $\lambda_2^{2IKM}$ are independent of impurity coupling $J'$. Our numerical fitting shows that $\lambda_1^{2IKM}=0.19$ and $\lambda_2^{2IKM}=0.46$ perfectly matches with the data. To see this, in Figs.~\ref{fig2}(c)-(d) we plot $E_1^{2IKM}(K_c)$ and $E_2^{2IKM}(K_c)$ as functions of $N^{0.19}$ and $N^{0.46}$ respectively for various values of $J'$ which all show perfect linear dependence. Although both $E_1$ and $E_2$ are defined in terms of bipartite entanglement quantities their behavior are completely different as for instance when $K\gg K_c$ both tripartite entanglements $E_1$ and $E_2$ vanish but the bipartite entanglement $N_{B,C}$ between the two bulks is significant due to an effective coupling $\sim J'^2/K$ \cite{bayat2012entanglement} induced by the impurities.

The same analysis can be done for the 2CKM in which for a fixed value of $J'$ we compute the tripartite entanglement as a function of asymmetry parameter $\Gamma$. The results are shown in Figs.~\ref{fig3}(a)-(b) for impurity coupling $J'=0.4$ and various system sizes. As the figures clearly show, the tripartite entanglement $E_1^{2CKM}$ and $E_2^{2CKM}$ peak at the critical point $\Gamma=\Gamma_c$ and its maximum value becomes even more pronounced by increasing the system size suggesting its divergence at the thermodynamic limit ($N{\rightarrow} \infty$).  Similar to before, by taking the values at criticality we find that  
\begin{align} \label{Emax_2CKM}
E_j^{2CKM}(\Gamma_c) \sim N^{\lambda_j^{2CKM}} \quad \text{(for } j=1,2), 
\end{align}
where our numerical fit results in $\lambda_1^{2CKM}=0.19$ and $\lambda_2^{2CKM}=0.5$.  In Figs.~\ref{fig3}(c)-(d) we plot $E_1^{2CKM}(\Gamma_c)$ and $E_2^{2CKM}(\Gamma_c)$ as functions of $N^{0.19}$ and $N^{0.5}$ respectively for various impurity couplings. the perfect linearity of the curves shows that the scaling of Eq.~(\ref{Emax_2CKM}) is very precise.  

All the above analysis suggest us to take the following ansatz for the tripartite entanglements for both 2IKM and 2CKM  
\begin{align} \label{Ej_Ansatz}
E_j=\frac{A}{|g-g_c|^{\beta_j} + B N^{-\lambda_j}} \quad \text{(for } j=1,2), 
\end{align}
where $g$ ($g_c$) should be replaced by $K$ ($K_c$) for 2IKM and $\Gamma$ ($\Gamma_c$) for 2CKM. 
The other two parameters, namely $A$ and $B$ are independent of $g$ and may only depend on $J'$. While the exponents  $\lambda_j$'s have been evaluated above the other exponents, i.e. $\beta_1$ and $\beta_2$, need more elaborate work and will be discussed in the following sections. 

\begin{figure}[t]
	\centering
	\includegraphics[width=\linewidth]{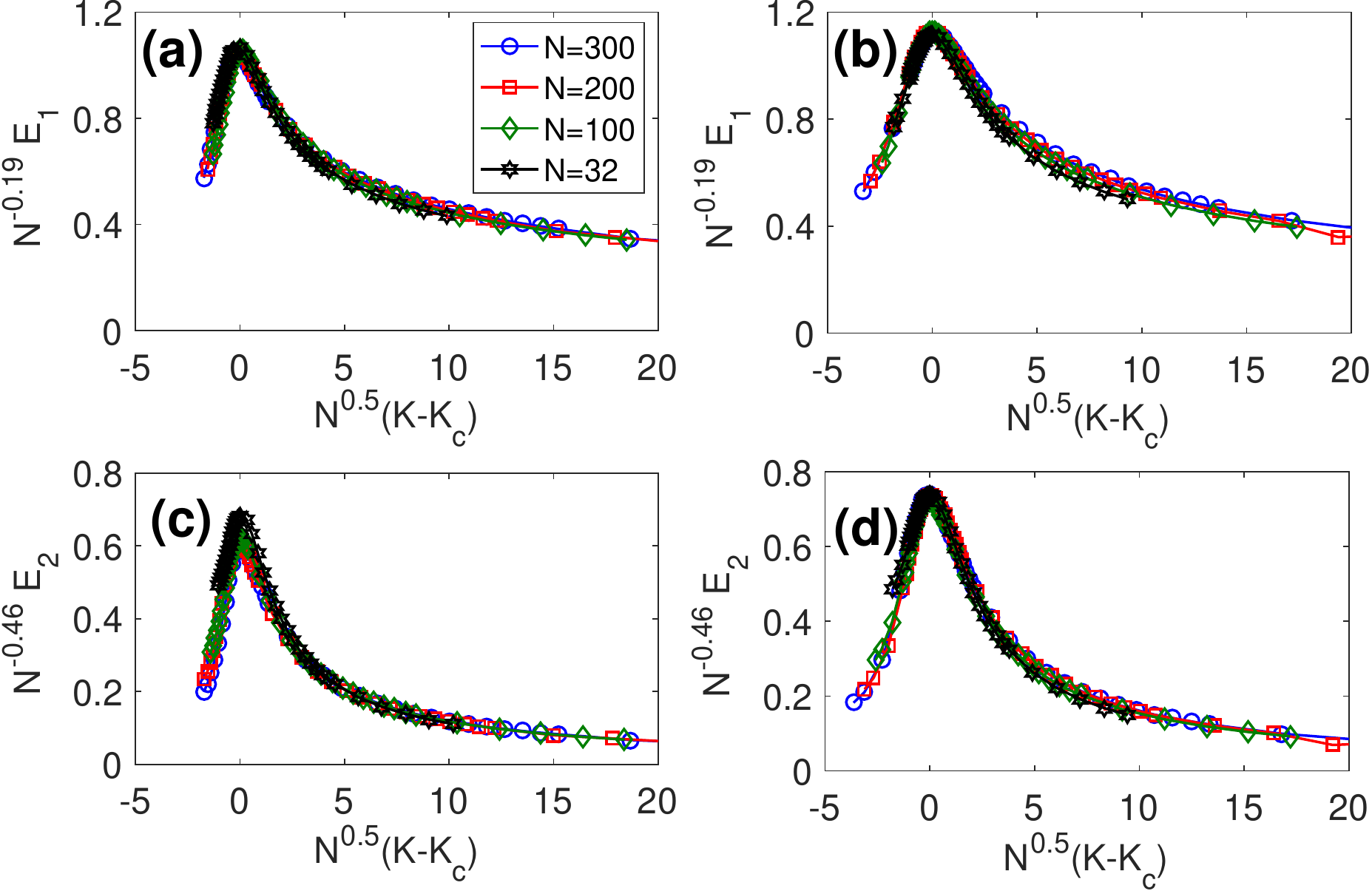}
	\caption{(color online) \textbf{Finite size scaling in 2IKM.} Data collapse of Eq.~(\ref{Finite_size}) for $E_1^{2IKM}$  in a chain with: (a) $J'=0.4$; and (b) $J'=0.5$. Similar data collapse for $E_2^{2IKM}$  in a chain with: (c) $J'=0.4$; and (d) $J'=0.5$. }
	\label{fig4}
\end{figure}

\emph{Scaling of tripartite entanglement.--}  A remarkable fact of QPTs is the emergence of a diverging length scale as $\xi\sim|g-g_c|^{-\nu}$ which results in scale invariant behavior for various quantities \cite{sachdev2007quantum}. 
To see if a complex many-body quantity such as tripartite entanglement also shows scaling we take a standard finite size ansatz as
\begin{align} \label{Finite_size}
E_j= N^{\beta_j/\nu} f(N^{1/\nu}|g-g_c|) \quad \text{(for } j=1,2),
\end{align}
where $f(\cdots)$ is a scaling function and $\beta_j$ is the same exponent as the one that appears in Eq.~(\ref{Ej_Ansatz}). In order to  evaluate the critical exponents $\nu$ and $\beta_j$ we search for those values of  $\nu$ and $\beta_j$'s such that the plots of $N^{-\beta_j/\nu}E_j$ as functions of $N^{1/\nu}|g-g_c|$ collapse on each other for various system sizes. We repeat this for both 2IKM and 2CKM separately. For the case of 2IKM the results for $E_1^{2IKM}$ are shown in Figs.~\ref{fig4}(a)-(b) for two impurity couplings $J'=0.4$ and $J'=0.5$ respectively. As these plots clearly show, a  very good data collapse can be achieved for both impurity couplings by choosing $\nu=2$ and $\beta_1^{2IKM}=0.38$. The same can be done for the second tripartite entanglement measure $E_2^{2IKM}$ and the results are shown in Figs.~\ref{fig4}(c)-(d)  for impurity couplings $J'=0.4$ and $J'=0.5$ respectively. As it can be seen from these figures the data collapse for $E_2$ can be achieved by $\nu=2$ and $\beta_2^{2IKM}=0.92$. The critical exponent $\nu=2$ is in perfect agreement with the results from conformal field theory \cite{affleck1995conformal} and Schmidt gap \cite{bayat2014order} analysis. 

\begin{figure}[t]
	\centering
	\includegraphics[width=\linewidth]{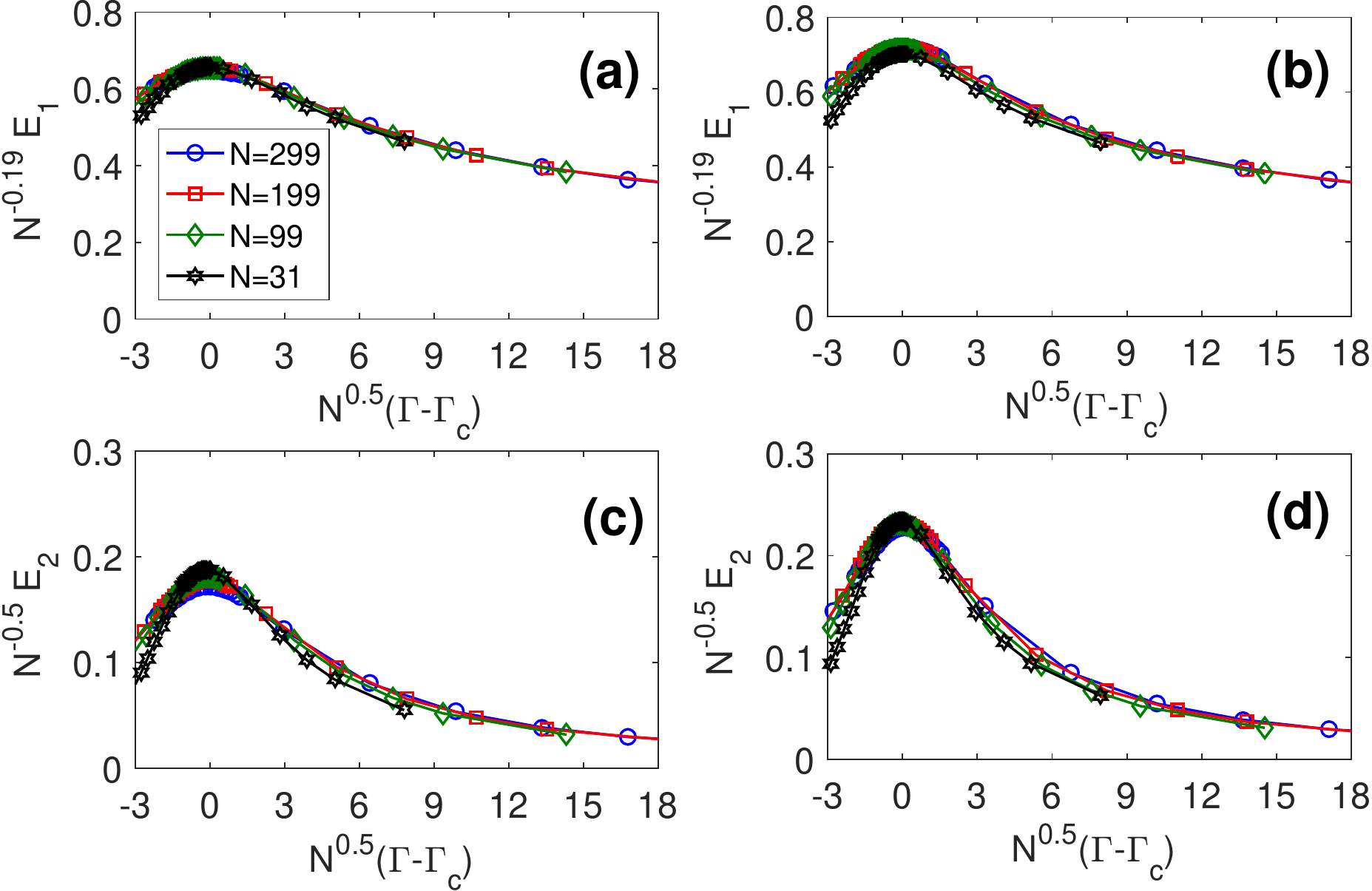}
	\caption{(color online) \textbf{Finite size scaling in 2CKM.} Data collapse of Eq.~(\ref{Finite_size}) for $E_1^{2CKM}$  in a chain with: (a) $J'=0.4$; and (b) $J'=0.5$. Similar data collapse for $E_2^{2CKM}$  in a chain with: (c) $J'=0.4$; and (d) $J'=0.5$.}
	\label{fig5}
\end{figure}

Similarly, for the 2CKM we can use the finite size scaling form of Eq.~(\ref{Finite_size}). The results for $E_1^{2CKM}$ are shown in Figs.~\ref{fig5}(a)-(b) for impurity couplings $J'=0.4$ and $J'=0.5$ respectively. The best data collapse are achieved by $\nu=2$, which is in full agreement with Ref.~\cite{alkurtass2016entanglement}, and $\beta_1^{2CKM}=0.38$. In Figs.~\ref{fig5}(c)-(d), for impurity couplings $J'=0.4$ and $J'=0.5$ respectively, we show the  the data collapse for  $E_2^{2CKM}$ is achieved by $\nu=2$ and $\beta_2^{2CKM}=1$. 

It is worth emphasizing that the critical exponent $\nu$, which shows how the length scale diverges near the critical point, is uniquely determined by the Hamiltonian of the system and is the same for all scaling quantities. Moreover, comparing the critical exponents $\beta_1^{2IKM}=0.38$ and $\beta_1^{2CKM}=0.38$ for our first tripartite entanglement measure (namely $E_1$) and $\beta_2^{2IKM}=0.92$ and $\beta_2^{2CKM}=1$ for the second tripartite entanglement measure (namely $E_2$) shows that the critical exponents are very close. 
This lends support to the idea that 2IKM and 2CKM belong to the same universality class \cite{mitchell2012two}.


\emph{Relationship between critical exponents.--} Comparing Eq.~(\ref{Ej_Ansatz}) with Eq.~(\ref{Finite_size}) may look that they are independent. However, by putting $g=g_c$ in Eq.~(\ref{Finite_size}) one can see that the critical exponents have to satisfy the following identity
\begin{align}
\beta_j=\nu \lambda_j.
\end{align}
The critical exponents for both measures $E_1$ and $E_2$ satisfy this identity meaning that only two of the three critical exponents are independent.

\emph{Conclusions.--} We have introduced two entanglement measures, based on negativity, for quantifying tripartite entanglement in impurity systems. While $E_1$ has already been proposed in Ref.~\cite{campbell2010multipartite}, the measure $E_2$ is new. Our analysis show that the tripartite entanglement, between impurities and the two bulks, in both 2IKM and 2CKM diverges at the critical point and shows scaling behavior with some critical exponents. Our analysis strongly suggests that the 2IKM and the 2CKM belong to the same universality class. 

\emph{Acknowledgements.--} Discussions with S. Bose,  H. Johannesson and P. Sodano are warmly acknowledged. This work has been supported by the EPSRC grant EP/K004077/1.


%

\end{document}